\documentclass[10pt, final, journal, letterpaper, twocolumn]{IEEEtran}

\makeatletter
\def\ps@headings{%
\def\@oddhead{\mbox{}\scriptsize\rightmark \hfil \thepage}%
\def\@evenhead{\scriptsize\thepage \hfil \leftmark\mbox{}}%
\def\@oddfoot{}%
\def\@evenfoot{}}
\makeatother 

\IEEEoverridecommandlockouts

\usepackage{amsfonts}
\usepackage[dvips]{graphicx}
\usepackage{caption}
\usepackage{subfigure}
\usepackage{times}
\usepackage{cite}
\usepackage{lettrine}
\usepackage{amsmath}
\usepackage{amsmath}
\allowdisplaybreaks[4]
\usepackage{array}
\usepackage{amssymb}

\usepackage{stfloats}
\usepackage{slashbox}
\usepackage{graphicx}
\usepackage{footnote}
\usepackage{booktabs}
\usepackage{array}
\usepackage{algorithmic}
\usepackage{algorithm}
\usepackage{subeqnarray}
\usepackage{cases}
\usepackage{threeparttable}
\usepackage{color}
\DeclareMathOperator*{\argmax}{argmax}

\newtheorem{proposition}{\underline{Proposition}}[section]

\begin{document}
\bibliographystyle{IEEEtran}

\title{Charge-then-Forward: Wireless Powered Communication for Multiuser Relay Networks}
\IEEEoverridecommandlockouts
\author{Mengyu Liu and Yuan Liu,~\IEEEmembership{Senior Member,~IEEE}

\thanks{

The authors are  with the School of Electronic and Information Engineering,
South China University of Technology, Guangzhou 510641, China (e-mail: liu.mengyu@mail.scut.edu.cn,  eeyliu@scut.edu.cn).
}
}

\maketitle

\vspace{-1.5cm}
\begin{abstract}

This paper studies a  relay-assisted wireless powered  communication network (R-WPCN) consisting of multiple source-destination pairs and a hybrid relay node (HRN).
We consider  a ``charge-then-forward" protocol at the HRN, in which the HRN with constant energy supply first acts as an energy transmitter to charge the sources, and then forwards the information from the sources to their destinations through time division multiple access (TDMA) or frequency division multiple access (FDMA).  Processing costs at the wireless-powered sources are taken into account.
Our goal is to maximize the sum-rate of all transmission pairs by jointly optimizing the time, frequency  and power resources.
The formulated optimization problems for both TDMA and FDMA are non-convex.
For the TDMA scheme, by  appropriate transformation, the problem is reformulated as a convex problem and be optimally solved.
For the FDMA case, we find the asymptotically optimal solution in the dual domain.
Furthermore, suboptimal algorithms are proposed for both schemes to tradeoff the complexity and performance.
Finally, the simulation results validate the effectiveness of the proposed schemes.

\end{abstract}



\section{Introduction}


The rapid growth of high-speed data and multimedia services  increases energy consumption for better quality-of-services. However, conventional battery-powered communications have to replace or recharge batteries manually to extend their lifetime, which is inconvenient, unsafe, and costly. Recently, radio-frequency (RF) signal enabled   wireless power transfer (WPT) has drawn great attention as it essentially provides more cost-effective and  green energy supplies for wireless devices, where RF signals are used as the carriers to convey wireless energy to low-power wireless devices.

There are two main directions of WPT among  the current related researches. One line of WPT focuses on so-called simultaneous wireless information and power transfer (SWIPT), where the  same RF signal carries both energy and information at the same time \cite{MIMO_zhangrui,SWIPT_morden}.
Due to the practical limitation of receivers that the received signals cannot be used to perform energy harvesting  and information decoding  simultaneously, two practical receiver  architectures, namely time switching (TS) and power splitting (PS), were proposed in \cite{MIMO_zhangrui}. For TS, the received signal is either used for energy harvesting or information decoding, whereas for PS, the received signal is split into two separate streams with one stream for energy harvesting and the other for information decoding at the same time.
SWIPT has been investigated extensively in different systems, e.g., the fading channels \cite{fading_channel}, relay channels \cite{relay_channel_2,relay_channel_3,12-relay,15-ts-relay} and orthogonal frequency division multiple access (OFDMA) channels \cite{OFDMA_channels_1,OFDMA_channels_2,OFDMA_channels_4,me,R1_x1}.

The newly emerging wireless powered communication network (WPCN) is another line of WPT where ambient RF signals are used to power wireless devices \cite{opportunitiesWPCN}.
There are two basic applications about WPCN.
One is that energy transmitters and information access points (APs) are located separately where energy transmitters transmit energy to wireless devices and then wireless devices transmit their information to APs using their harvested energy from energy transmitters \cite{7-WPCN}.
Another application is that a hybrid AP (HAP) performs the roles of energy transmitter and AP integrally.
For instance, a ``harvest-then-transmit'' protocol was proposed in \cite{1-wpcn-he}, where HAP first broadcasts wireless energy to all wireless devices in the downlink and then wireless devices utilize the harvested energy to transmit their independent information to HAP in the uplink based on TDMA.
Different from the HAP in SWIPT  that coordinates wireless energy and information, the HAP in WPCN only broadcasts wireless energy.
%

An important application for WPCN lies in relay-assisted  WPCN (R-WPCN),  where relays are used to assist information transmission in R-WPCN.
There are two categories among the current related works about R-WPCN: one is source powering relay \cite{R1_x2,4-user-co-relay,3-relay-full} and the second is relay powering source \cite{R3_x1,kwan_relaycapacity }.
%
For the first category, i.e., source powering relay,
the authors in \cite{R1_x2} proposed a ``harvest-then-cooperate" protocol, where both source and relay can harvest energy from the RF signals from a base-station.
A  two-user R-WPCN was studied in \cite{4-user-co-relay}  where a nearer user to HAP harvests energy sent by HAP and relays information of the farther user in half-duplex.
In \cite{3-relay-full}, the full-duplex relay not only is powered by the source but also harvests energy from itself by energy recycling.
%
As for the category of relay powering source,  the throughput maximization problem was investigated in \cite{R3_x1}, where the source can  harvest energy from the access point and/or relay before information transmission.
The authors in \cite{kwan_relaycapacity} studied the channel capacity subject to an additional energy transmission cost at the energy harvesting sources.
Note that both \cite{R3_x1} and \cite{kwan_relaycapacity} considered a single source-destination pair.

\begin{figure}[t]
\begin{centering}
\centering
\includegraphics[scale = 0.3]{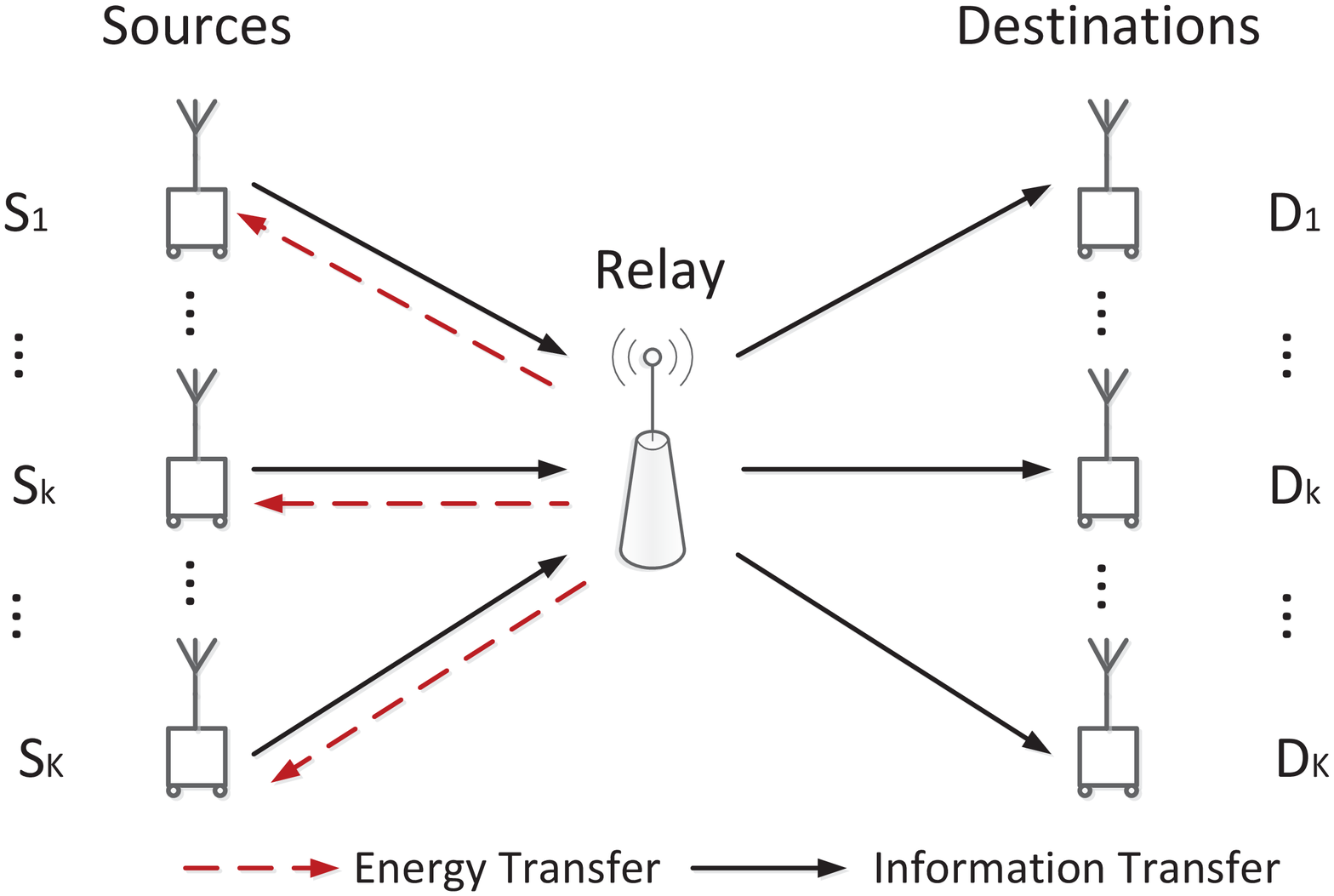}
\vspace{-0.1cm}
\caption{System model of the considered multiuser R-WPCN. }\label{fig:ff1}
\end{centering}
\vspace{-0.1cm}
\end{figure}
%

In this paper, we consider a new R-WPCN consisting of multiple source-destination pairs assisted by a single hybrid relay node (HRN), as shown in Fig. \ref{fig:ff1}.  We assume that the HRN in this paper has constant energy supply, while the sources nodes have no embedded power supply so that they have to be powered by the HRN before information transmission, i.e., ``harvest-then-transmit" protocol is applied at the sources. The HRN thus acts double roles, one for an energy transmitter and the other for an information helper. That is, the HRN first charges the sources and then forwards their information, i.e., ``charge-then-forward" protocol is considered at the HRN.

%
As the considered HRN has double roles, i.e., energy transmitter and information helper, energy charging and information forwarding of the HRN are mutually influenced and restricted since the HRN's total energy is fixed. That is, encouraging the energy charging will increase the transmit power of sources at the first hop but decrease the information forwarding at the second hop. How to find the optimal tradeoff that maximizes the system sum-rate is non-trivial. In addition, based on TDMA and FDMA for multiuser information transmission, the network resources, like time, power, and frequency, are highly coupled and the formulated optimization problems are non-convex and difficult to solve. Moreover, we consider the processing cost at the wireless-powered  sources, which further complicates the problems.

The main contributions of this paper are summarized as follows:
\begin{itemize}
  \item  We consider a new multiuser R-WPCN based on a ``charge-then-forward" relaying protocol, where the HRN first powers the energy-free sources and then forwards the information from the sources to their destinations by TDMA and  FDMA.  Processing cost is considered at the wireless-powered sources.

  \item  Depending on whether TDMA or FDMA  is adopted, we formulate two optimization problems respectively for sum-rate maximization, which are both non-convex. We propose efficient algorithms to find the  optimal solutions. In addition, suboptimal algorithms are proposed for both schemes to tradeoff the complexity and performance.

  \item  We provide some useful insights into the R-WPCN. For example,  the time of WPT should be as small as possible so that the time for wireless information transmission (WIT) can be maximized for sum-rate maximization. In addition,  due to the doubly distance-dependent signal attenuation for both WPT and the first hop of WIT, it is shown that the sum-rate decreases when the HRN moves from the sources to the destinations.

\end{itemize}

%

%

The remainder of this paper is organized as follows. In Section II, we introduce the system model  of multiuser R-WPCN and problem formulations based on TDMA and FDMA, respectively.
Section III presents the optimal and suboptimal resource allocation algorithms for the TDMA based problem. In  the next, the  asymptotically optimal and suboptimal algorithms for the FDMA based problem are presented in Section IV.
In Section V, we evaluate the performance of proposed algorithms by simulations. Finally, Section VI concludes the paper.

\section{System Model and Problem Formulation}

As shown in Fig. \ref{fig:ff1}, we consider a general two-hop R-WPCN with multiple source-destination pairs as well as a HRN which not only transfers energy to the sources but also forwards information from the sources to the destinations.  All nodes are equipped with a single antenna. We assume that the HRN is half-duplex due to the practical consideration, and there is no direct link between each source-destination pair due to the shielding effect caused by obstacles. As a result, each pair needs the assistance  of the HRN to forward information.
In this paper, we consider that the HRN has a constant energy supply while the source nodes have no embedded energy and thus have to harvest energy for information transmission. In addition, we assume that each source has the energy harvesting function to store the energy.
%
%
In particular, we consider the ``charge-then-forward" relaying protocol to coordinate power and information transfer, in which the HRN first acts as a wireless power beacon to charge the sources then as a helper for forwarding their information.
Specifically, the whole transmission is divided into two continuous phases. The first phase is used for WPT conducted by the HRN. The second phase is WIT, i.e.,  the sources use the harvested energy to transmit their independent information to their destinations via the assistance of the HRN in the second phase based on TDMA or FDMA. The sources do not store the harvested energy for future, i.e., all the energy harvested during WPT phase is used for WIT.


The global channel state information (CSI) of the network is assumed to be known at the HRN where the central processing task is embedded.
RF power transfer crucially depends on the available CSI of the nodes, which needs additional resources to acquire and the straightforward way is channel estimation via pilot signals, similar to conventional wireless communication systems. In our R-WPCN, whether the sources transmit pilots and the HRN estimates CSI, or the HRN transmits pilots and the sources estimate CSI, the sources are required to have initial energy to transmit/decode the pilot signals at the beginning of the training phase (before WPT in transmission phase). Thus it is reasonable to assume that the wireless-powered sources reserve some circuit power at the beginning for channel estimation, since the energy used to channel estimation is much smaller than that of information transmission in practice. CSI acquisition in WPT systems is very important but seems to be beyond of the scope of this paper.
In this paper, we consider a block fading wireless environment so that the channel impulse response can be treated as time invariant in the block duration. As a result, the channel gains within the block duration remain unchanged  (but can vary in different block durations).
For convenience, we assume that the transmission time of each block is normalized to be unit.

\subsection{TDMA Case}

 \begin{figure}
  \centering
  \subfigure[TDMA]{
    \label{fa} 
    \includegraphics[width=0.45\textwidth]{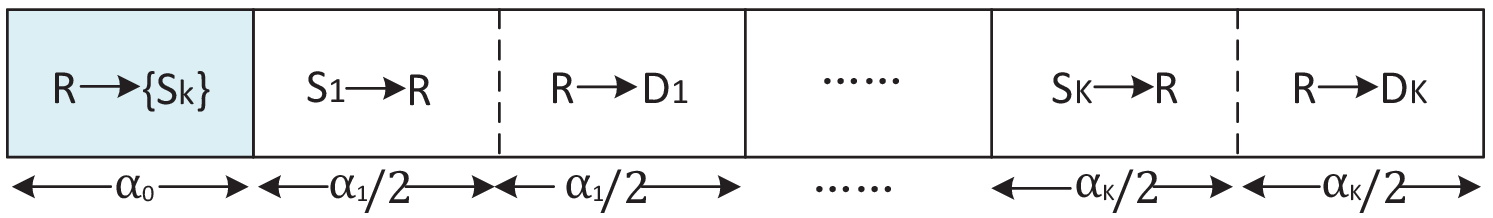}}
  \hspace{1in}
  \subfigure[FDMA]{
    \label{fb} 
    \includegraphics[width=0.45\textwidth]{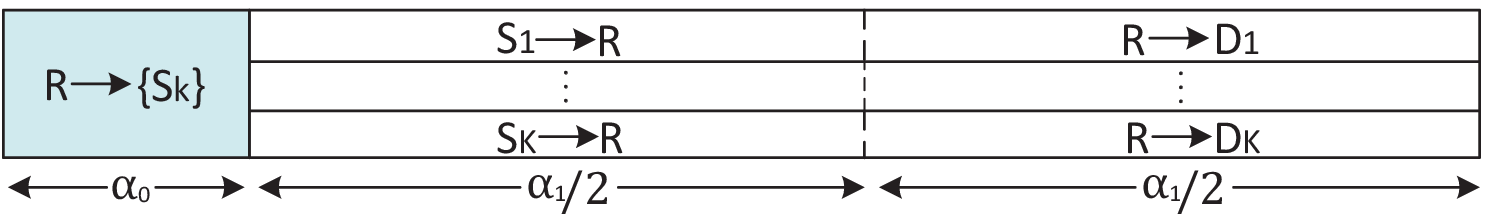}}
  \caption{The ``charge-then-forward" relaying protocol based on (a) TDMA and (b) FDMA information transmission.}
  \label{fig:subfig} 
\end{figure}

We first consider the case of TDMA-based information transmission as shown in Fig. 2(a). The total transmission time is divided into $K+1$ time slots where the first time slot, say slot 0 with time duration $\alpha_0$, is allocated for WPT and all source nodes harvest energy from the HRN,  while the rest $K$ slots are assigned to WIT of the $K$ pairs where each pair $k$ is allocated $\alpha_k$ time duration. Moreover, for each pair's information transmission, $\alpha_{k}$ is further divided into two equal sub-slots with $\alpha_k/2$ for the first hop and the rest $\alpha_k/2$ for the second hop.
By normalizing the whole time to be unit, we have
\begin{equation}
\sum\limits_{k=0}^K\alpha_{k}\leq1,\quad 0\leq\alpha_{k}\leq1, \quad k=0,1,\cdots,K.  \label{alpha_k_TDMA}
\end{equation}
In addition, the power of the HRN used at slot $k$ is denoted as $p_{k}$. Besides, we consider that there is a peak power constraint on $p_k$, i.e., $0 \leq p_k \leq P_{\rm{peak}}.$ Denote the maximum transmit power of the HRN as $P$,
%
then the energy constraint at the HRN is given by
\begin{equation}
\alpha_{0}p_{0}+\sum\limits_{k=1}^K\frac{\alpha_{k}}{2}p_k\leq P,  \label{power_C_TDMA}
\end{equation}
where  $\alpha_{k}/2$ is the transmission time of the HRN in the second hop for each pair $k$.  Note that the terms of power and energy are interchangeably used here since the duration of each block is normalized to be $T=1$ unit.

We consider energy accumulation for TDMA case, i.e., source $k$ harvests and accumulates energy from the previous slots, i.e., slot 0 to slot $k-1$.
The channel power gain from HRN to source $k$ for WPT and the channel power gain from source $i$ to source $k$ are denoted by $g_{{\rm r},k}$ and $g_{i,k}$, respectively.
Denote source $i$'s transmit power at slot $i$ as $q_i$.
Then the harvested energy of source $k$ can be expressed as
\begin{align}  \label{E_k_TDMA}
E_k = \left\{
        \begin{array}{ll}
          \eta\alpha_0p_0g_{{\rm r},k}, &\hbox{$k=1$} \\
          \eta \bigg( \alpha_0 p_0 g_{{\rm r},k} + \sum\limits_{i=1}^{k-1} \frac{\alpha_i }{2}  p_i g_{{\rm r},k} \\ \quad\quad\quad\quad + \sum\limits_{i=1}^{k-1} \frac{\alpha_i }{2} q_i g_{i,k}  \bigg), &\hbox{$k=2,\cdots,K,$}
        \end{array}
      \right.
\end{align}
%
which comprises three parts: the first term is the energy harvested in WPT phase, and the last two terms correspond to the energy harvested from the HRN and sources in the previous WIT phase. Here $0<\eta<1$ is the energy conversion efficiency at the sources.
%
%
As a result, the energy causality constraint at source $k$ is given by
\begin{align}
\frac{\alpha_k}{2} q_k + E_k^c \leq  E_k, \quad  k=1,\cdots,K, \label{C_energysource_TDMA}
\end{align}
where $\alpha_k/2$ is the time of the first hop for the information transmission of pair $k$ and $E_k^c$ is the non-zero energy processing cost at source $k$.
%
%
Moreover, the additional White Gaussian noise (AWGN) at  each  node is modeled as circularly symmetric complex Gaussian (CSCG) random variable with zero mean and variance $\sigma^{2}$. Denote the channel power gains for the first and second hops of WIT for pair $k$ as $h_{1,k}$ and $h_{2,k}$, respectively. Using decode-and-forward (DF) relaying strategy, the achievable rate for each pair $k$ is given by
\begin{align}  \label{R_k_TDMA}
R_{k}=
\frac{\alpha_{k}}{2}\min\left\{\log_{2}\left(1+\frac{q_k h_{1,k}}{\sigma^{2}}\right),\log_{2}\left(1+\frac{p_{k} h_{2,k}}{\sigma^{2}}\right)\right\}.
\end{align}

Our objective is to maximize the sum-rate of all pairs by jointly optimizing the time allocation, the transmit power of sources and HRN. Let  $\boldsymbol{p} = \{p_k\} $, $\boldsymbol{q} = \{q_k\} $,
$\boldsymbol{\alpha} = \{\alpha_k\} $ and $\boldsymbol{R} = \{R_k\} $,
the problem can be mathematically formulated as
\begin{subequations}
\begin{align}
{\rm (P1):}
\max_{\{\boldsymbol{p},\boldsymbol{q},\boldsymbol{\alpha},\boldsymbol{R}\}}\quad\quad&\sum\limits_{k=1}^K R_{k} \label{TDMA_c1}  \\
{\rm s.t.}\quad\quad&\eqref{alpha_k_TDMA}, \eqref{power_C_TDMA}, \eqref{C_energysource_TDMA},\eqref{R_k_TDMA}\nonumber \\
\quad\quad&0\leq p_{k}\leq P_{\rm{peak}},  k=0,1,\cdots,K.  \label{tdma_peak_c}
\end{align}
\end{subequations}

Problem (P1) is non-convex since the rate expression \eqref{R_k_TDMA} is not jointly concave in the variables. We will optimally solve this problem in Section III.

\subsection{FDMA Case}

We also consider the case of FDMA-based information transmission for the multiple pairs as shown in Fig. 2(b), where the total time is divided into two time slots, i.e., slot 0 and slot 1 utilized for WPT (using energy signals) and WIT, respectively.
The time duration of slot 0 and slot 1 are denoted by $\alpha_{0}$ and $\alpha_{1}$ with
\begin{equation}
\alpha_{0}+\alpha_{1}\leq1,\quad 0\leq \alpha_{0},\alpha_1\leq 1.  \label{x_k,n_OFDMA}
\end{equation}
We assume that the HRN broadcasts energy signals over the entire bandwidth in the phase of WPT, while information signals are conveyed by using FDMA over $N$ subcarriers (SCs) in the next phase of WIT.
For information transmission, we define a binary SC allocation variable $x_{k,n}$ with $x_{k,n}=1$ representing that SC $n$ is allocated to pair $k$ for WIT and $x_{k,n}=0$ otherwise. Each SC is allocated  to at most one pair at slot 1 for WIT to avoid interference.   The  constraint can be expressed as
\begin{eqnarray}
\sum\limits_{k=1}^Kx_{k,n}\leq 1, \forall n,\quad x_{k,n} \in \{0,1\},\quad \forall n, k=1,\cdots,K. \label{x_k,n}
\end{eqnarray}
The channel power gains for WPT of source $k$, the first and second hops of pair $k$ over SC $n$ for WIT are denoted as $g_{{\rm r},k}$, $h_{1,k,n}$ and $h_{2,k,n}$, respectively.
The transmit power of HRN for WPT at slot 0 is denoted as $p_{0}$, and the power of the HRN for forwarding pair $k$'s information on SC $n$ at slot 1 is $p_{k,n}$.
Note that WPT is conducted over the entire bandwidth, thus there is no index $n$ for $g_{{\rm r},k}$ and $p_0$.
The total transmit energy constraint of the HRN is thus given by
\begin{equation}
\alpha_{0}p_{0}+\sum\limits_{n=1}^N\sum\limits_{k=1}^K\frac{\alpha_{1}}{2} p_{k,n} \leq P,  \label{FDMA_P}
\end{equation}
where $\alpha_{1}/2$ represents the transmission time of  the HRN in the second hop.
%
%
%

Moreover, we define  source $k$'s transmit power on SC $n$ at slot 1 as $q_{k,n}$. Different from TDMA case, since all sources transmit their information at the same time in slot 1, the harvested energy of sources are only from HRN during WPT phase. Therefore, the energy  constraint at source $k$ is given by
%
\begin{equation}
\sum\limits_{n=1}^N\frac{\alpha_{1}}{2} q_{k,n} + E_k^c \leq \eta \alpha_0 p_0 g_{{\rm r},k},  \quad  k=1,\cdots,K,  \label{energy_C_OFDMA}
\end{equation}
where the $\alpha_{1}/2$ represents the time of the first hop during information transmission.

The achievable rates of the first and second hops for pair $k$ over SC $n$ can be respectively written as:
\begin{align}
R_{1,k,n} &= \frac{\alpha_{1}}{2N}\log_{2}\left(1+\frac{q_{k,n}h_{1,k,n}}{\sigma^{2}}\right),\forall n,k,\\
R_{2,k,n} &= \frac{\alpha_{1}}{2N} \log_{2}\left(1+\frac{p_{k,n}h_{2,k,n}}{\sigma^{2}}\right), \forall n,k.
\end{align}
The achievable rate of pair $k$ by using DF relaying strategy is the minimum of the rates achieved in the two hops, which can be expressed as
\begin{eqnarray} \label{FDMA_minrate}
R_{k,n} = \min\left\{ R_{1,k,n} ,R_{2,k,n}\right\},\quad \forall n,k=1,\cdots,K.
\end{eqnarray}

Our goal is  maximizing the sum-rate of all transmission pairs by varying the transmit power of the sources and HRN,  SC assignment and time allocation.
Let $\boldsymbol{p} = \{p_0,p_{k,n}\}$, $\boldsymbol{q}= \{q_{k,n}\}$, $\boldsymbol{x}=\{x_{k,n}\}$, $\boldsymbol{\alpha}=\{\alpha_0,\alpha_1\}$ and $\boldsymbol{R}=\{R_{k,n}\}$,  the optimization problem can be mathematically formulated as
\begin{subequations}
\begin{align}
&{\rm (P2):}
\max_{\{\boldsymbol{p},\boldsymbol{q},\boldsymbol{x},\boldsymbol{\alpha},\boldsymbol{R}\}}\quad\sum\limits_{k=1}^K\sum\limits_{n=1}^Nx_{k,n}R_{k,n} \label{P2}\\
&{\rm s.t.}\quad\eqref{x_k,n_OFDMA}, \eqref{x_k,n}, \eqref{FDMA_P}, \eqref{energy_C_OFDMA}\nonumber\\
&\quad\quad R_{k,n}\leq R_{1,k,n},R_{k,n}\leq R_{2,k,n}, \forall n,k=1\cdots,K,\label{r_k,n_10}\\
& \quad\quad 0\leq p_{0}, p_{k,n}\leq P_{\rm{peak}},  \forall n, k=1,\cdots,K. \label{peak_FDMA}
\end{align}
\end{subequations}

Problem (P2) is also non-convex  since both binary and continuous variables are involved, which is a mixed-integer programming problem. The asymptotically optimal solution for Problem (P2) will be obtained in Section IV.

\section{Resource Allocation in TDMA Case}


%

In this section, we study the TDMA case by solving Problem (P1). Problem (P1) is not convex and cannot be solved in its original form. Therefore, to make the problem tractable, we introduce a set of new variables $\boldsymbol{m} = \{\alpha_k q_k /2\}$ and $\boldsymbol{s}=\{\alpha_0p_0,\alpha_k p_k /2\}$. Clearly, $\boldsymbol{m}$ and $\boldsymbol{s}$ can be viewed as the actual transmit energy of the sources and HRN, respectively.
%
Problem (P1) is equivalent to the following problem:
\begin{subequations}
\begin{align}
&{\rm (P1'):} \label{TDMA_problem}
\max_{\{\boldsymbol{\alpha},\boldsymbol{m},\boldsymbol{s},\boldsymbol{R}\}}\quad\sum_{k=1}^K R_k  \\
&{\rm s.t.}\sum\limits_{k=0}^K\alpha_{k}\leq1,\quad 0\leq\alpha_{k}\leq1, \quad k=0,1,\cdots,K,  \label{alpha_s_TDMA}\\
&\quad \sum_{k=0}^{K} s_k \leq P, \label{energy_Q}\\
&\quad R_{k} \leq R_{1,k},  R_{k} \leq R_{2,k}, \quad  k=1,\cdots,K,\label{C_rate2_TDMA}\\
& \quad m_k + E_k^c \leq  \eta s_0 g_{{\rm r},k}, \quad k=1, \label{energy_source_TDMAk1}\\
& \quad m_k + E_k^c \leq  \eta \left( \sum_{i=0}^{k-1} s_i g_{{\rm r},k} + \sum_{i=1}^{k-1} m_i g_{i,k}  \right), k=2,\cdots,K, \label{energy_source_TDMA} \\
& \quad 0 \leq s_0 \leq \alpha_0 P_{\rm peak},  0 \leq s_k \leq \frac{\alpha_k}{2} P_{\rm peak},  k=1,\cdots,K. \label{alpha_k_peak}
\end{align}
\end{subequations}
where
\begin{align}
R_{1,k} &= \frac{\alpha_k}{2} \log_2 \left(1 + \frac{2 m_k h_{1,k}}{\alpha_k \sigma^2} \right),\quad  k=1,\cdots,K,\\
R_{2,k} &= \frac{\alpha_k}{2} \log_2 \left(1 + \frac{2 s_k h_{2,k}}{\alpha_k \sigma^2} \right),\quad  k=1,\cdots,K.
\end{align}


Since constraint \eqref{C_rate2_TDMA} is convex and the other constraints of Problem (P1$'$) are affine, Problem (P1$'$) is convex in its current form.
In the literature \cite{first_order_1,first_order_2,first_order_3,first_order_4}, the first-order method can be used to solve these non-convex problems by approximating the non-convex objective functions and constraints into convex ones. However, in this paper, by appropriate variable transformation, Problem (P1) is reformulated to be convex, which can thus be optimally solved by applying the Lagrange duality method, as will be shown next.

We first introduce non-negative Lagrangian multipliers $\boldsymbol{\lambda} = \{\lambda_k\}\succeq0$ and $\boldsymbol{\beta}=\{\beta_k\}\succeq 0$ associated with the rate constraint \eqref{C_rate2_TDMA}, $\boldsymbol{\nu}=\{\nu_k\}\succeq0$ associated with the energy  causality constraints \eqref{energy_source_TDMAk1} and \eqref{energy_source_TDMA}. In addition, non-negative Lagrangian multipliers $\mu\geq 0$ and $\xi\geq 0$ are associated with the total time constraint \eqref{alpha_s_TDMA} and total energy constraint at HRN \eqref{energy_Q}.
Then, the Lagrangian of Problem (P1$'$) is given by
\begin{align}
&\mathcal{L}\left(\boldsymbol{s},\boldsymbol{m},\boldsymbol{\alpha},\boldsymbol{R},\boldsymbol{\lambda},\boldsymbol{\beta},\boldsymbol{\nu},\mu,\xi\right) \nonumber \\
= & \sum_{k=1}^{K} \left[ R_k +   \lambda_k \left( R_{1,k} - R_k \right) +  \beta_k \left( R_{2,k} - R_k \right)\right]\nonumber \\
& +\mu \left( 1-\sum_{k=0}^{K} \alpha_k \right) + \xi \left( P - \sum_{k=0}^{K} s_k \right) \nonumber \\
 & + \nu_1(\eta s_0 g_{{\rm r},1}-E_1^c-m_1) \nonumber \\
 & + \sum_{i=2}^{K}\nu_i \left(\sum_{k=0}^{i-1} \eta s_k g_{{\rm r},i} + \sum_{k=1}^{i-1} \eta m_k g_{k,i} - E_i^c - m_i \right). \label{L_TDMA_R}
\end{align}
Denote $\mathcal{D}$ as the set of $\{\boldsymbol{s},\boldsymbol{m},\boldsymbol{\alpha},\boldsymbol{R}\}$ satisfying the primary constraints, then the dual function of Problem (P1$'$) is given by
\begin{align}
g(\boldsymbol{\lambda},\boldsymbol{\beta},\boldsymbol{\nu},\mu,\xi) = \max_{\{\boldsymbol{s},\boldsymbol{m},\boldsymbol{\alpha},\boldsymbol{R}\} \in \mathcal{D}} \mathcal{L}\left(\boldsymbol{s},\boldsymbol{m},\boldsymbol{\alpha},\boldsymbol{R},\boldsymbol{\lambda},\boldsymbol{\beta},\boldsymbol{\nu},\mu,\xi\right). \label{dual_function_TDMA}
\end{align}
To compute the dual function $g(\boldsymbol{\lambda},\boldsymbol{\beta},\boldsymbol{\nu},\mu,\xi)$, we need to find the optimal $\{\boldsymbol{s}^*,\boldsymbol{m}^*,\boldsymbol{\alpha}^*,\boldsymbol{R}^*\}$ to maximize the Lagrangian under the
given dual variables $\{\boldsymbol{\lambda},\boldsymbol{\beta},\boldsymbol{\nu},\mu,\xi\}$.  In the following we present the derivations in detail.

\subsection{Optimizing $\{\boldsymbol{s},\boldsymbol{m},\boldsymbol{\alpha},\boldsymbol{R}\}$ for Given  $\{\boldsymbol{\lambda},\boldsymbol{\beta},\boldsymbol{\nu},\mu,\xi\}$}

\subsubsection{Maximizing Lagrangian over $\{R_{k}\}$} The part of the dual function with respect to the rate variable $\{R_{k}\}$ is given by
\begin{align}
g_R(\boldsymbol{\lambda},\boldsymbol{\beta}) = \max_{\boldsymbol{R}\succeq 0} \sum_{k=1}^{K} (1-\lambda_k-\beta_k)R_k.
\end{align}
To make sure that the dual function is bounded, we have $1-\lambda_k-\mu_k \equiv 0$. In such case, $g_R(\boldsymbol{\lambda},\boldsymbol{\beta})\equiv0$ \cite{Mojianhua} and we obtain that $\beta_k = 1-\lambda_k$. Note that $0\leq \lambda_k \leq 1$ such that $\beta_k$ is non-negative. By substituting these results above into \eqref{L_TDMA_R}, the Lagrangian can be rewritten as:
\begin{align}
&\mathcal{L}\left(\boldsymbol{s},\boldsymbol{m},\boldsymbol{\alpha},\boldsymbol{\lambda},\boldsymbol{\nu},\mu,\xi\right) \nonumber \\
= & \sum_{k=1}^{K}  [\lambda_kR_{1,k} + (1-\lambda_{k})R_{2,k}]+ \mu \left( 1-\sum_{k=0}^{K} \alpha_k \right) \nonumber \\
&+ \xi \left( P - \sum_{k=0}^{K} s_k \right)  + \nu_1(\eta s_0 g_{{\rm r},1}-E_1^c-m_1) \nonumber \\
&+ \sum_{i=2}^{K}\nu_i \left(\sum_{k=0}^{i-1} \eta s_k g_{{\rm r},i} + \sum_{k=1}^{i-1} \eta m_k g_{k,i} - E_i^c - m_i \right). \label{L_TDMA_NoR}
\end{align}

\subsubsection{Maximizing Lagrangian over $\{m_k\}_{k=1}^{K},$ $\{s_k\}_{k=1}^{K}$ and $\{\alpha_k\}_{k=1}^{K}$} Observing the Lagrangian in \eqref{L_TDMA_NoR}, we find that the dual function in \eqref{dual_function_TDMA} can be decomposed into $K+1$ independent functions:
\begin{flalign}
g(\boldsymbol{\lambda},\boldsymbol{\nu},\mu,\xi) = \sum_{k=0}^{K} g_k (\boldsymbol{\lambda},\boldsymbol{\nu},\mu,\xi) + \mu + \xi P - \sum_{i=1}^{K} \nu_i E_i^c, \label{dual_zuhe}
\end{flalign}
where
\begin{align} \label{g_k_TDMA}
g_k (\boldsymbol{\lambda},\boldsymbol{\nu},\mu,\xi) \triangleq \max_{\{\boldsymbol{s},\boldsymbol{m},\boldsymbol{\alpha}\} \in \mathcal{D}}
\mathcal{L}_k \left(\boldsymbol{s},\boldsymbol{m},\boldsymbol{\alpha},\boldsymbol{\lambda},\boldsymbol{\nu},\mu,\xi\right)
\end{align}
with
\begin{align} \label{L_k_TDMA}
& \mathcal{L}_k \left(\boldsymbol{s},\boldsymbol{m},\boldsymbol{\alpha},\boldsymbol{\lambda},\boldsymbol{\nu},\mu,\xi\right) = \nonumber\\
& \left\{
  \begin{array}{ll}
    -\mu_k \alpha_k - \xi s_k + \sum\limits_{i=1}^{K} \eta \nu_i g_{{\rm r},i} s_k,  \quad\quad\quad\quad\quad \quad \quad \quad \hbox{$k=0$,} \\
\lambda_k R_{1,k} + (1-\lambda_k) R_{2,k} \\ - \mu \alpha_k - \xi s_k - \nu_k m_k \\ + \sum\limits_{i=k+1}^{K} \eta \nu_i ( g_{{\rm r},i} s_k +  g_{k,i} m_k) , \quad\quad\quad \hbox{$k=1,\cdots,K-1$,} \\
    \lambda_k R_{1,k} + (1-\lambda_k) R_{2,k} - \mu \alpha_k - \xi s_k - \nu_k m_k, \quad \hbox{$k=K$.}
  \end{array}
\right.
\end{align}

%

For given dual point $\{\boldsymbol{\lambda},\boldsymbol{\nu},\mu,\xi\}$, maximizing \eqref{L_TDMA_NoR} over $\{m_k\}_{k=1}^{K},\{s_k\}_{k=1}^{K},\{\alpha_k\}_{k=1}^{K}$ is equivalent to solving  \eqref{g_k_TDMA} for $k=1,\cdots,K.$  From \eqref{L_k_TDMA}, the partial derivatives of $\mathcal{L}_k$ with respect to $s_k$ and $m_k$ can be given by \eqref{piandaos} and \eqref{piandaom} on the top of the next page.
\begin{figure*}[ht]
\begin{align}
\frac{\partial \mathcal{L}_k}{\partial s_k} &=\left\{
                                               \begin{array}{ll}
                                                 \frac{(1-\lambda_k)\alpha_k h_{2,k}}{(\alpha_k \sigma^2 + 2s_k h_{2,k})\ln2} + \sum\limits_{i=k+1}^{K}\eta \nu_i g_{{\rm r},i} - \xi, & \hbox{$k=1,\cdots,K-1$} \\
                                                 \frac{(1-\lambda_k)\alpha_k h_{2,k}}{(\alpha_k \sigma^2 + 2s_k h_{2,k})\ln2}  - \xi, & \hbox{$k=K$}
                                               \end{array}
                                             \right. \label{piandaos}\\
\frac{\partial \mathcal{L}_k}{\partial m_k} &= \left\{
                                                \begin{array}{ll}
                                                  \frac{\lambda_k\alpha_k h_{1,k}}{(\alpha_k \sigma^2 + 2m_k h_{1,k})\ln2} + \sum\limits_{i=k+1}^{K}\eta \nu_i g_{k,i} - \nu_k, & \hbox{$k=1,\cdots,K-1$} \\
                                                  \frac{\lambda_k\alpha_k h_{1,k}}{(\alpha_k \sigma^2 + 2m_k h_{1,k})\ln2} - \nu_k, & \hbox{$k=K$}
                                                \end{array}
                                              \right. \label{piandaom}
\end{align}
\hrulefill
\end{figure*}
Given $\alpha_k,k=1,\cdots,K$, the optimal energy variables $s_k$ and $m_k$ that maximize $\mathcal{L}_k$ can be obtained by setting $\frac{\partial \mathcal{L}_k}{\partial s_k}=0$ and $\frac{\partial \mathcal{L}_k}{\partial m_k} =0$ and are given by \eqref{s_k_TDMA} and \eqref{m_k_TDMA} on the top of the next page.
\begin{figure*}[ht]
\begin{align}
s_k &= \left\{
         \begin{array}{ll}
            \frac{\alpha_k}{2} \min \left\{ \left[ \frac{1-\lambda_k}{\left(\xi - \sum_{i=k+1}^{K} \eta \nu_i g_{{\rm r},i}\right)\ln2} - \frac{\sigma^2}{h_{2,k}} \right]^+,P_{\rm peak} \right\}, & \hbox{$k=1,\cdots,K-1$} \label{s_k_TDMA} \\
           \frac{\alpha_k}{2} \min \left\{ \left( \frac{1-\lambda_k}{\xi \ln2} - \frac{\sigma^2}{h_{2,k}} \right)^+,P_{\rm peak} \right\}, & \hbox{$k=K$}
         \end{array}
       \right.\\
m_k &= \left\{
         \begin{array}{ll}
           \frac{\alpha_k}{2} \left[ \frac{\lambda_k}{\left( \nu_k - \sum_{i=k+1}^{K} \eta\nu_ig_{k,i} \right)\ln2} - \frac{\sigma^2}{h_{1,k}}, \right]^+, & \hbox{$k=1,\cdots,K-1$}  \label{m_k_TDMA}\\
           \frac{\alpha_k}{2} \left( \frac{\lambda_k}{ \nu_k \ln2} - \frac{\sigma^2}{h_{1,k}}, \right)^+, & \hbox{$k=K$}
         \end{array}
       \right.
\end{align}
\hrulefill
\end{figure*}

%
With given $s_k$ and $m_k$,
we can easily prove that $\frac{\partial \mathcal{L}_k}{\partial \alpha_k}$ is a decreasing function of $\alpha_k$. As a result, the optimal $\alpha_k$ with given $s_k$ and $m_k$ can be found by a simple bisection search over $0\leq \alpha_k \leq 1.$

To summarize, for $k=1,\cdots,K$, Problem \eqref{g_k_TDMA} can be solved by iteratively optimizing between $\{s_k,m_k\}$ and $\alpha_k$ with one of them fixed at one time, which is known as block-coordinate descent (BCD) method.

\subsubsection{Maximizing Lagrangian over $s_0$ and $\alpha_0$} Next, we study the solution of Problem \eqref{g_k_TDMA} for $k=0$, which is a linear programming problem (LP). From \eqref{L_k_TDMA}, to maximize $\mathcal{L}_0$ we have
\begin{align}
s_0 &= \left\{
        \begin{array}{ll}
          \alpha_0 P_{\rm peak}, & \hbox{if $-\xi + \sum\limits_{i=0}^{K} \eta \nu_i g_{{\rm r},i} > 0$,} \label{s0_TDMA}\\
          0, & \hbox{otherwise.}
        \end{array}
      \right. \\
\alpha_0 &= \left\{
        \begin{array}{ll}
          1, & \hbox{if $-\mu -\xi P_{\rm peak} + \sum\limits_{i=0}^{K} \eta \nu_i g_{{\rm r},i} P_{\rm peak}> 0$,}  \label{alpha0_TDMA}\\
          0, & \hbox{otherwise.}
        \end{array}
      \right.
\end{align}

\subsection{Optimizing Dual Variables $\{\boldsymbol{\lambda},\boldsymbol{\nu},\mu,\xi\}$}

As a dual function is always convex \cite{convexbook}, we adopt the ellipsoid method to simultaneously iterate the dual variables $\{\boldsymbol{\lambda},\boldsymbol{\nu},\mu,\xi\}$ to the optimal ones by using the defined subgradients as follows:


\begin{equation}     \label{citidu}  
\boldsymbol{\triangle}=\left[                 
  \begin{array}{lcl}   
    \Delta \lambda_{k} = R_{1,k}-R_{2,k}, k=1,\cdots,K\\  
    \Delta \mu  = 1-\sum\limits_{k=0}^{K}\alpha_k\\  
\Delta \xi =P-\sum\limits_{k=0}^{K}s_k\\
\Delta \nu_1 = \eta s_0 g_{{\rm r},1} - E_1^c - m_1\\
\Delta \nu_k = \sum\limits_{i=0}^{k-1}\eta s_i g_{{\rm r},k} - \sum\limits_{i=1}^{k-1}\eta m_i g_{i,k}\\ \quad\quad\quad- E_k^c - m_k, k=2,\cdots,K
  \end{array}
\right]                 
\end{equation}

\subsection{Discussion on Optimality and Complexity}

The optimal $s_k^*$, $m_k^*$ and $\alpha_k^*$ for $k=1,\cdots,K$ are obtained at optimal $\{\boldsymbol{\lambda}^*,\boldsymbol{\nu}^*,\mu^*,\xi^*\}$, then the optimal $\alpha_0^*$ is given by $\alpha_0^* = 1-\sum_{k=1}^{K}\alpha_k^*$.  With $\{\alpha_k^*\}_{k=0}^{K}$, $\{s_k^*\}_{k=1}^{K}$ and $\{m_k^*\}_{k=1}^{K}$, Problem (P1$'$) becomes a LP with variable $s_0$. The optimal value of $s_0^*$ is obtained by solving this LP.


To summarize, the algorithm to solve Problem (P1$'$) is given in Algorithm 1. The time complexity of steps 3-7 is of order $K^2$. The complexity of step 9 is $\mathcal{O}(K^2)$. Therefore, the complexity of steps 3-9 is given by $\mathcal{O}(K^2)$. Note that step 10 iterates $\mathcal{O}(q^2)$ to converge, where $q$ is the number of dual variables and $q=2K+2$ in our case.
Thus the complexity of steps 1-10 is $\mathcal{O}(q^2K^2)$. The time complexity of the LP is $\mathcal{O}(K)$. Therefore, the complexity of Algorithm 1 is $\mathcal{O}(q^2K^2+K)$.

\begin{algorithm}[tb]
\caption{Optimal Algorithm for Problem (P1$'$)}
\begin{algorithmic}[1]
\STATE Initialize $\{\boldsymbol{\lambda},\boldsymbol{\nu},\mu,\xi\}$.
\REPEAT
\STATE Initialize $\alpha_k = 1/K, k = 1,\cdots,K$.
\REPEAT
\STATE Compute $\{s_k\}_{k=1}^{K}$ and $\{m_k\}_{k=1}^{K}$ by \eqref{s_k_TDMA} and \eqref{m_k_TDMA}, respectively.
\STATE Obtain $\{\alpha_k\}_{k=1}^{K}$  with given $\{s_k\}$ and $\{m_k\}$ by bisection search.
\UNTIL improvement of $\mathcal{L}_k,k=1,\cdots,K$ converges to a prescribed accuracy.
\STATE Compute $s_0$ and $\alpha_0$ by \eqref{s0_TDMA} and \eqref{alpha0_TDMA}, respectively.
\STATE Update $\{\boldsymbol{\lambda},\boldsymbol{\nu},\mu,\xi\}$ according to the ellipsoid method via \eqref{citidu}.
\UNTIL $\{\boldsymbol{\lambda},\boldsymbol{\nu},\mu,\xi\}$ converge to a prescribed accuracy.
\STATE Set $s_k^*=s_k$, $m_k^*=m_k$,  $\alpha_k^*=\alpha_k$ for $k=1,\cdots,K$, and $\alpha_0^* = 1-\sum\limits_{k=1}^{K}\alpha_k^*$.
\STATE Obtain $s_0^*$ by solving Problem (P1$'$) with $\{s_k^*\}_{k=1}^{K}$, $\{m_k^*\}_{k=1}^{K}$ and$\{\alpha_k^*\}_{k=0}^{K}$.
\end{algorithmic}
\end{algorithm}

\begin{proposition} \label{TDMA_peak_infty}
For the TDMA case with $K$ source-destination pairs and $P_{\rm peak}\to\infty$, the maximum sum-rate by solving Problem (P1$'$) is  achieved by $\alpha_0^*\to0$.
\end{proposition}

\textit{Proof:} Clearly, we have $\alpha_0>0$ and $s_0>0$; otherwise, no energy will be harvested at the sources. Since the objective function of Problem (P1$'$) is an increasing function of $\alpha_k$ for $k=1,\cdots,K$ from constraint \eqref{C_rate2_TDMA},  when it comes to the extreme case with $P_{\rm peak}\to\infty$, for any given $s_k$ and $m_k$ satisfying constraints \eqref{energy_Q}, \eqref{energy_source_TDMAk1} and \eqref{energy_source_TDMA}, the optimal solution must be achieved by $\sum_{k=1}^{K}\alpha_k\to 1$  according to constraint \eqref{alpha_s_TDMA}. In this case, $\alpha_0^*\to0$ and $p_0^*\to \infty$ are required to guarantee positive harvested energy at the sources. The proof is thus completed.

\begin{proposition} \label{TDMA_peak_fty}
For the TDMA case with $K$ source-destination pairs and finite $P_{\rm{peak}}$,  the maximum sum-rate for Problem (P1$'$) is achieved by $p_{0}^*=P_{\rm{peak}}$.
\end{proposition}

\textit{Proof:} Please refer to Appendix A.

By Proposition \ref{TDMA_peak_infty} and Proposition \ref{TDMA_peak_fty}, it can be inferred that Problem (P1$'$) is actually a problem of energy and time allocation at the HRN, i.e., allocating energy and time for WPT and each WIT. Therefore,  for any given energy  allocated for WPT (i.e., $s_0=\alpha_0 p_0$), the HRN should charge the sources at its maximum available power (i.e., $p_0=P_{\rm peak}$), so that the time used for WPT $\alpha_0=s_0/p_0$ can be as small as possible and more time $1-\alpha_0$ can be allocated to WIT due to the  sum-rate maximization goal. In particular, when $P_{\rm peak}\to \infty$, the portion of transmission time $\alpha_0$ for WPT should asymptotically go to zero, which means that the sources can harvest sufficient energy in a sufficiently small time and  almost whole time is allocated to  WIT.

\subsection{Suboptimal Algorithm}

The complexity of the optimal algorithm becomes high as the number of pairs increases, mainly due to the dual updates. By simplifying the system model and eliminating the dual updates, in this section, we present an efficient suboptimal algorithm which significantly reduces the complexity.

At first, in WIT phase, the received power at each source in other periods is from the relay and other sources, which are both small. Specifically, the received energy from other sources is negligible due to the double energy
decay, i.e., the energy decay of relay-to-source and then source-to-source. As DF relaying protocol is adopted, the transmission power of the relay could relatively match the source's transmit power, and thus the relay's transmit power for forwarding is also small. As a result, in this section, we consider that the harvested energy at the sources is only from the WPT phase. With give $\alpha_0$, the transmit power of source $k$ can be given by
\begin{align}
q_k = \frac{2(\eta\alpha_0p_0g_{{\rm r},k}-E_k^c)^+}{\alpha_k},\quad  k=1,\cdots,K. \label{q_k+}
\end{align}

Second, due to Proposition \ref{TDMA_peak_fty}, we let $p_0=P_{\rm peak}$. Moreover, we assume that the equal power allocation (EPA) at the HRN in the WIT phase, the transmit power at the HRN for pair $k$ is thus given by
\begin{align} \label{average_P_TDMA}
p_k = \min\left\{\frac{2(P-\alpha_0P_{\rm peak})}{1-\alpha_0},P_{\rm peak}\right\}, k =1,\cdots,K.
\end{align}

Third, due to the energy decay in the WPT phase, the transmit power of sources may be small, thus the performance of this considered dual-hop relaying system may depend on the rate of first hop under most cases. As a result, in this section, we only focus on maximizing the sum rate of the first hop. Therefore, we have the following problem:
\begin{align}
\max_{\boldsymbol{\alpha}\in \mathcal{D}}\quad\quad&\sum\limits_{k=1}^K \frac{\alpha_k}{2}\log_2\left(1+\frac{2A_k }{\alpha_k\sigma^2}\right) \label{sumrate_first_TDMA}
\end{align}
where $A_k \triangleq (\eta\alpha_0P_{\rm peak}g_{{\rm r},k}-E_k^c)^+h_{1,k}$.

\begin{proposition} \label{sub_TDMA}
The optimal solution of Problem \eqref{sumrate_first_TDMA} with given $\alpha_0$ is given by
\begin{align}
q_k &= \frac{2}{(1-\alpha_0)h_{1,k}}\sum_{k=1}^{K} A_k,\quad  k=1\cdots,K,\label{q_sub_TDMA}\\
\alpha_k &= \frac{ A_k}{\sum_{k=1}^{K} A_k } (1-\alpha_0),\quad  k=1\cdots,K.\label{alpha_sub_TDMA}
\end{align}
\end{proposition}

\textit{Proof: }Please refer to Appendix B.

With given $\alpha_0$, we can obtain a set of $\{\boldsymbol{\alpha},\boldsymbol{p},\boldsymbol{q}\}$ by \eqref{average_P_TDMA}, \eqref{q_sub_TDMA} and \eqref{alpha_sub_TDMA}. Then, the optimal $\alpha_0$ maximizing the sum-rate can be found by the one-dimensional search.
%
\begin{algorithm}[tb]
\caption{Suboptimal Algorithm for Problem (P1)}
\begin{algorithmic}[1]
\STATE Divide $\alpha_{0}$ in $[0,1]$ with fixed step $\epsilon$.
\FOR{ each $\alpha_0 P_{\rm peak} \leq  P$ }
\STATE Compute the time allocation for WIT $\{\alpha_k\}_{k=1}^{K}$ according to \eqref{alpha_sub_TDMA}.
\STATE Compute the power allocation for WIT $\{p_k\}_{k=1}^{K}$ and $\{q_k\}_{k=1}^{K}$ according to  \eqref{average_P_TDMA} and \eqref{q_sub_TDMA}, respectively.
\STATE Compute the sum-rate according to \eqref{R_k_TDMA} with given $\alpha_0$.
\ENDFOR
\STATE Choose the optimal $\alpha_0^*$ that has the maximum sum-rate.
\end{algorithmic}
\end{algorithm}

To summarize, the above suboptimal algorithm is given in Algorithm 2.  The complexity of steps 3-5 is $\mathcal{O}(K)$. The complexity for searching $\alpha_0$ is $\mathcal{O}(1/\epsilon)$. Therefore, the whole complexity of Algorithm 2 is $\mathcal{O}(K/\epsilon)$, which is  linear in $K$ and much lower than that of the optimal algorithm in above subsection.

\section{Resource Allocation in FDMA Case}

Problem (P2) is a mixed integer programming and thus is NP-hard and non-convex.  However, it has been shown that the duality gap of the resource allocation problems in FDMA systems becomes zero when the number of SCs goes to large \cite{dual_gap1,dual_gap2}. This means that the optimal solution obtained in dual domain is equivalent to the optimal solution of the original non-convex problem due to the zero duality gap. Thus we solve Problem (P2) in dual domain.

At first, we introduce non-negative Lagrangian multipliers $\boldsymbol{\lambda}=\{\lambda_{k,n}\}\succeq 0$ and $\boldsymbol{\beta}=\{\beta_{k,n}\}\succeq 0$ corresponding to the two rates of the first and second hops in \eqref{r_k,n_10}, and $\boldsymbol{\nu}=\{\nu_k\}\succeq 0$ associated with the energy causality constraint \eqref{energy_C_OFDMA}. Moreover, $\mu\geq 0$, $\xi\geq 0$ are introduced to associate with the total time constraint \eqref{x_k,n_OFDMA} and total energy constraint \eqref{FDMA_P}, respectively. Then the dual function of Problem (P2) can be defined as
\begin{align}
g(\boldsymbol{\lambda},\boldsymbol{\beta},\boldsymbol{\nu},\mu,\xi) \triangleq \max_{\{\boldsymbol{p},\boldsymbol{q},\boldsymbol{x},\boldsymbol{\alpha},\boldsymbol{R}\}\in\mathcal{D}} \mathcal{L}(\boldsymbol{p},\boldsymbol{q},\boldsymbol{x},\boldsymbol{\alpha},\boldsymbol{R},\boldsymbol{\lambda},\boldsymbol{\beta},\boldsymbol{\nu},\mu,\xi),
\end{align}
where $\mathcal{D}$ is the set of all primal variables $\{\boldsymbol{p},\boldsymbol{q},\boldsymbol{x},\boldsymbol{\alpha},\boldsymbol{R}\}$ satisfying  the constraints, and the Lagrangian of Problem (P2) is
\begin{align} \label{L_FDMA}
&\mathcal{L}(\boldsymbol{p},\boldsymbol{q},\boldsymbol{x},\boldsymbol{\alpha},\boldsymbol{R},\boldsymbol{\lambda},\boldsymbol{\beta},\boldsymbol{\nu},\mu,\xi) \nonumber\\
= &\sum_{k=1}^{K}\sum_{n=1}^{N} x_{k,n} [ R_{k,n}+ \lambda_{k,n}(R_{1,k,n}-R_{k,n}) \nonumber\\
&+ \beta_{k,n}(R_{2,k,n}-R_{k,n}) ] + \mu (1-\alpha_0-\alpha_1) \nonumber\\
&+ \xi\left( P -\alpha_0 p_0 - \sum_{n=1}^{N} \sum_{k=1}^{K} \frac{\alpha_1}{2}p_{k,n}\right) \nonumber\\
&+ \sum_{k=1}^{K} \nu_k \left(\eta \alpha_0 p_0 g_{{\rm r},k}  - \sum_{n=1}^{N} \frac{\alpha_1}{2}q_{k,n}- E_k^c\right).
\end{align}

Computing the dual function $g(\boldsymbol{\lambda},\boldsymbol{\beta},\boldsymbol{\nu},\mu,\xi)$ requires to determine the optimal $\{\boldsymbol{p},\boldsymbol{q},\boldsymbol{x},\boldsymbol{\alpha},\boldsymbol{R}\}$ for given dual variables $\{\boldsymbol{\lambda},\boldsymbol{\beta},\boldsymbol{\nu},\mu,\xi\}$. In the following we present the derivations in detail.

\subsection{Optimizing $\{\boldsymbol{p},\boldsymbol{q},\boldsymbol{x},\boldsymbol{\alpha},\boldsymbol{R}\}$ for Given  $\{\boldsymbol{\lambda},\boldsymbol{\beta},\boldsymbol{\nu},\mu,\xi\}$}

\subsubsection{Maximizing Lagrangian over $\{R_{k,n}\}$} Similar to TDMA case, the part of dual function with respect to $\{R_{k,n}\}$ is given by
\begin{align}
g_R(\boldsymbol{\lambda},\boldsymbol{\beta},\boldsymbol{\nu},\mu,\xi) = \max_{\boldsymbol{R}\succeq 0} \sum_{k=1}^{K} \sum_{n=1}^{N} (1-\lambda_{k,n}-\beta_{k,n})x_{k,n}R_{k,n}.
\end{align}

To make sure that the dual function is bounded, we have $1-\lambda_{k,n}-\beta_{k,n} \equiv 0$. In such case, $g_R(\boldsymbol{\lambda},\boldsymbol{\beta},\boldsymbol{\nu},\mu,\xi) \equiv 0$ and we obtain that $\beta_{k,n} = 1 - \lambda_{k,n}.$
Note that $0\leq\lambda_{k,n}\leq1$ to make sure that $\beta_{k,n}$ is non-negative. By substituting the result above into \eqref{L_FDMA}, we have
\begin{align} \label{L_FDMA_xiao}
&\mathcal{L}(\boldsymbol{p},\boldsymbol{q},\boldsymbol{x},\boldsymbol{\alpha},\boldsymbol{\lambda},\boldsymbol{\nu},\mu,\xi) \nonumber\\
=&  \sum_{k=1}^{K} \sum_{n=1}^{N} x_{k,n} \left[ \lambda_{k,n}R_{1,k,n} + (1-\lambda_{k,n})R_{2,k,n} \right] \nonumber\\
&+ \mu (1-\alpha_0-\alpha_1) \nonumber\\
&+ \xi\left( P -\alpha_0 p_0 - \sum_{n=1}^{N} \sum_{k=1}^{K} \frac{\alpha_1}{2}p_{k,n}\right) \nonumber\\
& + \sum_{k=1}^{K} \nu_k \left(\eta \alpha_0 p_0 g_{{\rm r},k}  - \sum_{n=1}^{N} \frac{\alpha_1}{2}q_{k,n}- E_k^c\right).
\end{align}

\subsubsection{Maximizing Lagrangian over $\{p_{k,n}\}$, $\{q_{k,n}\}$, $\{x_{k,n}\}$ and $\alpha_1$ } Observing the Lagrangian in \eqref{L_FDMA_xiao}, we can rewrite \eqref{L_FDMA_xiao} as follows:
\begin{align}
&\mathcal{L}(\boldsymbol{p},\boldsymbol{q},\boldsymbol{x},\boldsymbol{\alpha},\boldsymbol{\lambda},\boldsymbol{\nu},\mu,\xi) \nonumber\\
=& \sum_{n=1}^{N}\mathcal{L}_n (\boldsymbol{p},\boldsymbol{q},\boldsymbol{x},\boldsymbol{\alpha},\boldsymbol{\lambda},\boldsymbol{\nu},\xi) + \mu(1-\alpha_0-\alpha_1) \nonumber\\
 & + \xi(P -\alpha_0p_0) + \sum_{k=1}^{K}\nu_k\left(\eta\alpha_0p_0g_{{\rm r},k}-E_k^c\right),
\end{align}
where
\begin{align} \label{Ln_FDMA}
&\mathcal{L}_n (\boldsymbol{p},\boldsymbol{q},\boldsymbol{x},\boldsymbol{\alpha},\boldsymbol{\lambda},\boldsymbol{\nu},\xi)\nonumber\\
=&\sum_{k=1}^{K} x_{k,n} \left[ \lambda_{k,n}R_{1,k,n} + (1-\lambda_{k,n})R_{2,k,n} \right]\nonumber\\
&- \sum_{k=1}^{K} \frac{\alpha_1}{2}\left( \xi p_{k,n} + \nu_k q_{k,n} \right).
\end{align}
Maximizing $\mathcal{L}$ over $\{p_{k,n}\}$ and $\{q_{k,n}\}$ is equivalent to maximizing each $\mathcal{L}_n$ over $p_{k,n}$ and $q_{k,n}$, which is shown in the following. Each SC $n$ should be allocated to at most one pair and we can apply exhaustive method for SC to obtain the optimal $k^*$. Particulary, assume that SC $n$ is selected for pair $k$, then \eqref{Ln_FDMA} is equivalent to
\begin{align}  \label{Ln_fix}
\mathcal{L}_n (p_{k,n},q_{k,n},\alpha_1,\lambda_{k,n},\nu_k,\xi)
=& \lambda_{k,n} R_{1,k,n} +  (1-\lambda_{k,n}) R_{2,k,n}\nonumber\\
-&  \frac{\alpha_1}{2}\left( \xi p_{k,n} + \nu_k q_{k,n} \right).
\end{align}
By differentiating \eqref{Ln_fix} with respect to $p_{k,n}$ and $q_{k,n}$, and letting them to zero, the $p_{k,n}$ and $q_{k,n}$ maximizing $\mathcal{L}_n$ are given by
\begin{align}
p_{k,n} &= \min\left\{\left( \frac{1-\lambda_{k,n}}{\xi N \ln2} - \frac{\sigma^2}{h_{2,k,n}}\right)^+,P_{\rm peak}\right\}, \label{p}\\
q_{k,n} &= \left( \frac{\lambda_{k,n}}{\nu_k N \ln2} - \frac{\sigma^2}{h_{1,k,n}}\right)^+.\label{q}
\end{align}

After computing the power allocations $p_{k,n}$ and $q_{k,n}$ for WIT, we can obtain the SC allocation  maximizing each $\mathcal{L}_n$ as
\begin{eqnarray}  \label{eq:x(k,n)}
x_{k,n}=
\begin{cases}
1,       &\mbox{if} \quad k=\mathop{\argmax}_{k}\mathcal{L}_{n},\\
0,   &\mbox{otherwise}.
\end{cases}
\end{eqnarray}
With given $p_{k,n}$, $q_{k,n}$ and $x_{k,n}$, the Lagrangian \eqref{L_FDMA_xiao} becomes a linear function of $\alpha_1$. From \eqref{L_FDMA_xiao}, to maximize $\mathcal{L}$, we have
\begin{align} \label{alpha1_FDMA}
\alpha_1 = \left\{
             \begin{array}{ll}
               1, & \hbox{if $\Psi  > 0$,} \\
               0, & \hbox{otherwise.}
             \end{array}
           \right.
\end{align}
where $\Psi$ is given by
\begin{align}
\Psi =& \sum_{k=1}^{K}\sum_{n=1}^{N} x_{k,n} \left[ \lambda_{k,n}R_{1,k,n} + (1-\lambda_{k,n})R_{2,k,n} \right] \nonumber\\
&-  \frac{1}{2}\sum_{k=1}^{K}\sum_{n=1}^{N} x_{k,n}(\xi p_{k,n} - \nu_kq_{k,n}).
\end{align}
Here, $p_{k,n}$, $q_{k,n}$ and $x_{k,n}$ have been obtained by \eqref{p}, \eqref{q} and \eqref{eq:x(k,n)}, respectively.

\subsubsection{Maximizing Lagrangian over $\alpha_0$ and $p_0$} From \eqref{L_FDMA_xiao}, to maximize $\mathcal{L}$, we have
\begin{align}
\alpha_0 &= \left\{
             \begin{array}{ll}
               1, & \hbox{if $\sum\limits_{k=1}^{K} \eta \nu_k g_{{\rm r},k} P_{\rm peak} -\xi P_{\rm peak} - \mu > 0$,} \\
               0, & \hbox{otherwise.}
             \end{array}
           \right.\label{alpha_0FDMA}\\
p_0 &= \left\{
        \begin{array}{ll}
          P_{\rm peak}, & \hbox{if $\sum\limits_{k=1}^{K} \eta \nu_k g_{{\rm r},k} - \xi > 0$,} \\
          0, & \hbox{otherwise.}
        \end{array}
      \right. \label{p0FDMA}
\end{align}

\subsection{Optimizing Dual Variables $\{\boldsymbol{\lambda},\boldsymbol{\nu},\mu,\xi\}$}

Similar to TDMA case, the ellipsoid method can be employed to update $\{\boldsymbol{\lambda},\boldsymbol{\nu},\mu,\xi\}$ toward optimal $\{\boldsymbol{\lambda}^*,\boldsymbol{\nu}^*,\mu^*,\xi^*\}$ with global convergence \cite{convexbook}, the subgradients required for which are
\begin{equation}     \label{subtidu}  
\boldsymbol{\triangle}=\left[                 
  \begin{array}{lcl}   
    \Delta \lambda_{k,n} = x_{k,n}(R_{1,k,n}-R_{2,k,n}) , \forall k,n\\  
    \Delta \mu  = 1-\alpha_0-\alpha_1\\  
\Delta \xi = P -\alpha_0p_0 - \sum\limits_{k=1}^{K}\sum\limits_{n=1}^{N}\frac{\alpha_1}{2}p_{k,n}\\
\Delta \nu_k = \eta\alpha_0p_0g_{{\rm r},k} - \sum\limits_{n=1}^{N}\frac{\alpha_1}{2}q_{k,n} - E_k^c, \forall k
  \end{array}
\right]                 
\end{equation}

\subsection{Discussions on Optimality and Complexity}

It is worth noting that the optimal $\boldsymbol{p}^*$, $\boldsymbol{q}^*$ and $\boldsymbol{x}^*$ are obtained at optimal $\{\boldsymbol{\lambda}^*,\boldsymbol{\nu}^*,\mu^*,\xi^*\}$. With given optimal $\boldsymbol{p}^*$, $\boldsymbol{q}^*$ and $\boldsymbol{x}^*$, Problem (P2) becomes a LP with $\boldsymbol{\alpha}$. The optimal $\boldsymbol{\alpha}^*$ can be obtained by solving this LP.

\begin{algorithm}[tb]
\caption{Optimal Algorithm for Problem (P2)}
\begin{algorithmic}[1]
\STATE \textbf{initialize} $\{\boldsymbol{\lambda},\boldsymbol{\nu},\mu,\xi\}$.
\REPEAT
\FOR{each SC $n$}
\STATE Compute  $\{p_{k,n}\}$  and $\{q_{k,n}\}$ according to  \eqref{p} and \eqref{q}, respectively.
\STATE Obtain the  subcarrier allocation $\{x_{k,n}\}$ according to \eqref{eq:x(k,n)}.
\ENDFOR
\STATE Compute $\alpha_1,\alpha_0,p_0$ according to \eqref{alpha1_FDMA}, \eqref{alpha_0FDMA} and \eqref{p0FDMA}, respectively.
\STATE Update $\{\boldsymbol{\lambda},\boldsymbol{\nu},\mu,\xi\}$  by the ellipsoid method  using the subgradients defined in \eqref{subtidu}.
\UNTIL $\{\boldsymbol{\lambda},\boldsymbol{\nu},\mu,\xi\}$ converge to a prescribed accuracy.
\STATE Set $\boldsymbol{p}^* = \boldsymbol{p}$, $\boldsymbol{q}^* = \boldsymbol{q}$ and $\boldsymbol{x}^*=\boldsymbol{x}$.
\STATE Obtain $\boldsymbol{\alpha}^*$ by solving Problem (P2) with $\boldsymbol{p} = \boldsymbol{p}^*$, $\boldsymbol{q} = \boldsymbol{q}^*$ and $\boldsymbol{x}=\boldsymbol{x}^*$.
\end{algorithmic}
\end{algorithm}

To summarize, the algorithm for Problem (P2) is given by Algorithm 3. The complexity of steps 4-5 is  $\mathcal{O}(K N)$. The complexity of steps 3-6 is $\mathcal{O}(K N^2)$.
The ellipsoid method needs complexity of $\mathcal{O}(q)$, where $q$ is the number of dual variables and $q=KN+K+2$ in our case. The time complexity of the LP is $\mathcal{O}(K)$. Therefore, the whole complexity of Algorithm 3 is $\mathcal{O}(q^2KN^2 + K)$.

\begin{proposition}\label{FDMA_0}
In the case of the FDMA case with $K$ source-destination pairs and $P_{\rm peak} \to \infty$, the  maximum sum-rate obtained by solving Problem (P2) is given by $\alpha_0^*\to 0$ and $\alpha_1^* \to 1$.
\end{proposition}

\textit{Proof:} Clearly, we can easily obtain that $\alpha_0>0$; otherwise, no energy is harvested at the sources  in the considered R-WPCN.  Thus,  $\alpha_1<1$.  Denote $s_{0}=\alpha_0 p_{0}$, $s_{k,n}=\alpha_1 p_{k,n}/2$ and $m_{k,n}=\alpha_1 q_{k,n}/2$ as energy variables. For ang given $s_0,s_{k,n}$, $m_{k,n}$ and $x_{k,n}$ satisfying the primary constraints \eqref{x_k,n_OFDMA}, \eqref{x_k,n}, \eqref{FDMA_P}, \eqref{energy_C_OFDMA}, the objective function of Problem (P2) is an increasing function of $\alpha_1$ according to \eqref{r_k,n_10}. Thus, the sum-rate is maximized when $\alpha_1^* \to 1$, which follows $\alpha_0^*\to 0$.  The proof is thus completed.

By Proposition \ref{FDMA_0},  the positive harvested energy at the sources is achieved under the assumption that the HRN is able to transmit an infinite power due to $\alpha_0 \to 0$. For a finite $P_{\rm peak}$, a nonzero time ratio should be scheduled to the WPT phase to harvest sufficient energy for WIT.
Similar to the TDMA case, when it comes to the more general case with $P_{\rm peak}<\infty$, we have the following proposition,

\begin{proposition} \label{FDMA_dingli_peak}
In the case of the FDMA case with $K$ source-destination pairs and $P_{\rm peak} < \infty$, the  maximum sum-rate is achieved by $p_0^*=P_{\rm peak}$.
\end{proposition}

\textit{Proof:} The proof is similar as the proof of Proposition \ref{TDMA_peak_fty}, and thus is omitted here.

\subsection{Suboptimal Algorithm}

The main complexity of the above optimal algorithm is resulted from the ellipsoid method. In this section, we propose a suboptimal algorithm by assuming equal power allocation (EPA) over SCs, which can eliminate the dual updates, while the optimal $\alpha_0$ is  obtained by the one-dimensional search.

First, due to the energy decay in the WPT phase, the transmit power of sources may be small, thus the performance of this considered dual-hop relaying system may depend on the rate of first hop under most cases. As a result, we heuristically  choose the pair having the maximum $h_{1,k,n}$ to occupy SC $n$,  which is given by
\begin{eqnarray}  \label{sub_x}
x_{k,n}=
\begin{cases}
1,       &\mbox{if} \quad k=k^{\ast}=\mathop{\argmax}_{k} h_{1,k,n},\\
0,   &\mbox{otherwise}.
\end{cases}
\end{eqnarray}

With given $\alpha_0$, we let $p_0=P_{\rm peak}$ according to Proposition \ref{FDMA_dingli_peak}. Then, from  \eqref{FDMA_P} and \eqref{energy_C_OFDMA}, the power allocation by assuming EPA can be given by
\begin{align}
p_{k,n} &=\left\{
  \begin{array}{ll}
     \min \left\{\left[ \frac{2(P-\alpha_0 P_{\rm peak})}{\alpha_1 N}\right ]^+,P_{\rm peak}\right\}, & \hbox{if $x_{k,n}=1$,} \label{p_sub}\\
    0, & \hbox{otherwise.}
  \end{array}
\right.\\
q_{k,n}&=\left\{
  \begin{array}{ll}
     \frac{2 (\eta\alpha_0 P_{\rm peak}g_{{\rm r},k} - E_k^c)^+ }{\alpha_1 M_k}, & \hbox{if $x_{k,n}=1$,} \label{q_sub} \\
    0, & \hbox{otherwise.}
  \end{array}
\right.
\end{align}
where $M_k$ is the number of SCs assigned to source $k$, which can be determined via \eqref{sub_x}.



\begin{algorithm}[tb]
\caption{Subptimal Algorithm for Problem (P2)}
\begin{algorithmic}[1]
\STATE Divide $\alpha_{0}$ in $[0,1]$ with fixed step $\epsilon$.
\FOR{ each $\alpha_{0}P_{\rm{peak}}\leq P $}
\STATE Obtain the SC allocation $\{x_{k,n}\}$ by \eqref{sub_x}.
\STATE Compute the power allocations $\{p_{k,n}\}$ and $\{q_{k,n}\}$ according to \eqref{p_sub} and \eqref{q_sub}, respectively.
\STATE Compute the sum-rate according to \eqref{FDMA_minrate} for given $\alpha_0$.
\ENDFOR
\STATE Choose the optimal $\alpha_0^*$ that has the maximum sum-rate.
\end{algorithmic}
\end{algorithm}

The above suboptimal algorithm is summarized in Algorithm 4. The complexity of  steps 3-5 is $\mathcal{O}(K N)$. The complexity for searching $\alpha_0$ is $\mathcal{O}(1/\epsilon)$,
therefore the whole complexity of the algorithm is $\mathcal{O}(K N/\epsilon)$, which is  much lower than that of Algorithm 3.

%

\section{Numerical Results}

\begin{figure}[t]
\begin{centering}
\centering
\includegraphics[scale=1.2]{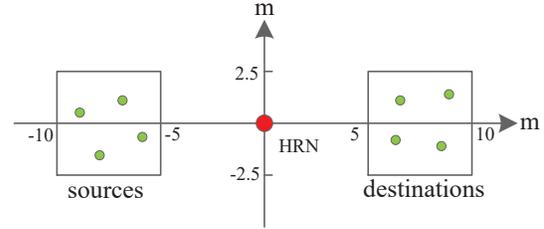}
\vspace{-0.1cm}
\caption{The location of R-WPCN in the simulations.}\label{location}
\end{centering}
\vspace{-0.1cm}
\end{figure}

In this section, we provide extensive numerical results to evaluate the performance of the proposed algorithms.
In the simulation setup, we assume that the total bandwidth is 10 MHz for both TDMA and FDMA cases and  the noise spectral density is assumed to be $-174$ dBm/Hz. We set  the energy conversion
efficiency as $\eta=0.8$  and the processing cost as $E_k^c = 1 \times 10^{-7}$ Joule   at all sources.
For all simulations, we set $K=4$ pairs of sources-destinations  unless otherwise noted.
Moreover,  we consider a two-dimensional plane of node location as shown in Fig. \ref{location}, where the source nodes and destination nodes are randomly but uniformly distributed in the corresponding square regions, and the HRN can move along with the x-axis from $-$5 m to 5 m. The HRN is assumed to locate at (0,0) unless otherwise noted.
%
%
In this paper, the pass-loss exponent is 3 and we adopt Richan fading channel model for the small-scale fading, where the Richan factor is set to be 3.

\begin{figure}[t]
\begin{centering}
\includegraphics[scale=0.65]{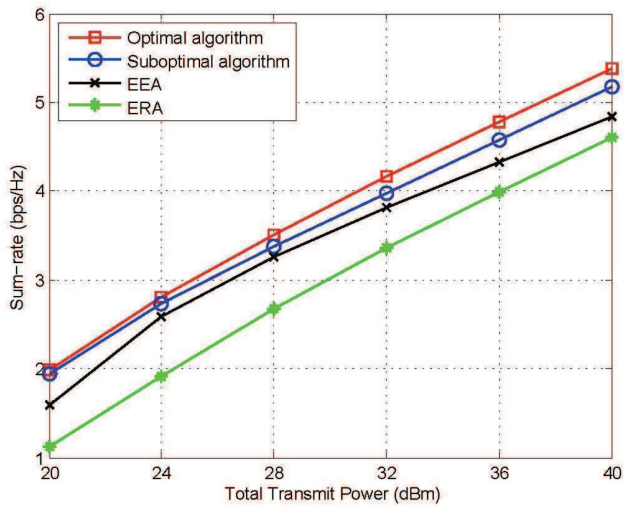}
\vspace{-0.1cm}
\caption{Sum-rate versus  total transmit power $P$ in  TDMA-based  R-WPCN, where $P_{\rm peak}=2P$.  }\label{fig:TDMA_performance_compare}
\end{centering}
\vspace{-0.1cm}
\end{figure}

Fig. \ref{fig:TDMA_performance_compare} demonstrates the sum-rate versus the total transmit power $P$ in the TDMA case with $P_{\rm peak}=2P$.
We introduce  the following two benchmark schemes for the purpose of performance comparison. First, the equal energy allocation (EEA) scheme is considered, where the energy allocated for WPT at the HRN is fixed as $\alpha_0 P_{\rm peak}=P/2$, while the optimal time allocation is still obtained as Algorithm 1.
In addition, we consider the equal resource allocation (ERA) for WIT with given $\alpha_0$, where the equal time and power allocations for WIT are assumed and the optimal $\alpha_0^*$ is obtained as Algorithm 2.
For all  schemes, the sum-rate is observed to increase with the total transmit power $P$.
Compared with EEA and ERA, we can see that the proposed optimal and suboptimal algorithms achieve  better performance.
And the suboptimal algorithm is observed to perform very closely to the optimal algorithm, which demonstrates the effectiveness of the proposed suboptimal resource allocations.
%

\begin{figure}
\begin{centering}
\includegraphics[scale=0.63]{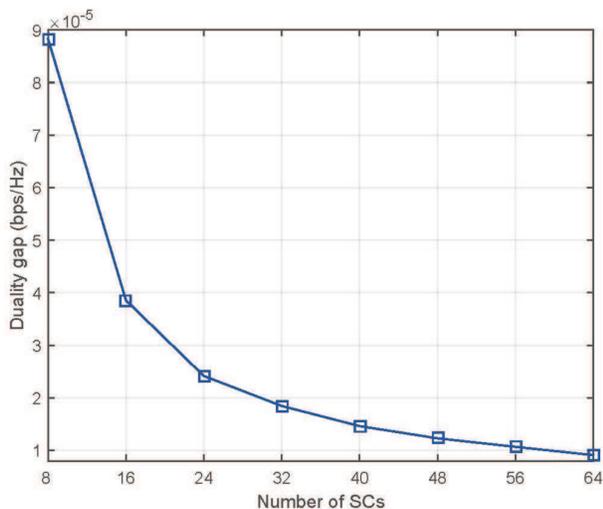}
\vspace{-0.1cm}
\caption{Duality gap versus number of SCs, where $P=30$ dBm, $P_{\rm peak}=2P$.} \label{fig:dual_gap}
\end{centering}
\vspace{-0.1cm}
\end{figure}

Fig. \ref{fig:dual_gap} illustrates the duality gaps versus the different numbers of SCs $N$. The duality gap is shown to decrease with the SCs number $N$. It can be observed that the duality gap is indeed approximately zero with 64 SCs, which  verifies the effectiveness of the proposed dual-based algorithm of FDMA case.

\begin{figure}[t]
\begin{centering}
\includegraphics[scale=0.65]{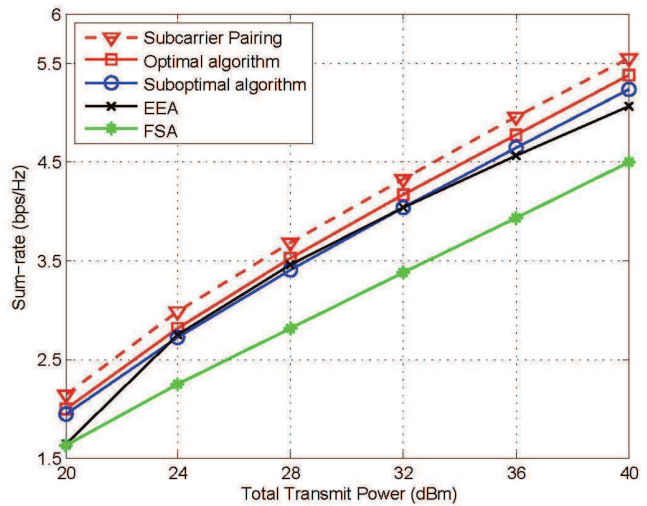}
\vspace{-0.1cm}
\caption{Sum-rate versus total transmit power $P$ in FDMA-based  R-WPCN, where $N=64$, $P_{\rm{peak}}=2P$.} \label{FDMA_perform}
\end{centering}
\vspace{-0.1cm}
\end{figure}

Fig. \ref{FDMA_perform} demonstrates the sum-rate versus the total transmit power $P$ in the FDMA case. The  peak power constraint $P_{\rm{peak}}$ is set to be $2P$.
For performance comparison, we consider the following three benchmarking schemes.
The first one is subcarrier pairing scheme where the subcarrier allocation in the two hops can be different \cite{pairing1}.
The second  one is equal energy allocation (EEA) where the  WPT energy  is fixed as $\alpha_0  P_{\rm peak} = P/2$, while the optimal power and SC allocations are still obtained as Algorithm 3.
The last benchmark is that the subcarrier assignment is fixed (FSA) while the optimal time and power allocations are also jointly optimized as Algorithm 3.
For all schemes, we can observe that the sum-rate is increasing with the total transmit power $P$ and the proposed optimal scheme achieves considerable gain compared with the other benchmark  schemes.
Compared with the optimal algorithm, it can be observed that the performance gain achieved by subcarrier pairing is limited but requires additional $\mathcal{O}(N^3)$ complexity by Hungarian algorithm.
Besides, the suboptimal algorithm with low complexity also has good performance.
Moreover, we can observe that the EEA scheme is only efficient under some particular system setup ($24 < P < 32$ dBm), while the suboptimal algorithm has a good performance over a wide range of transmit power compared with the EEA scheme, which demonstrates the superiority  of the proposed suboptimal algorithm.
The poor performance of FSA compared to the proposed schemes  indicates that dynamic SC allocation  provides significant improvement in terms of the sum-rate.



 两种模型下对比  峰值功率为横坐标
\begin{figure}[t]
\begin{centering}
\includegraphics[scale=0.65]{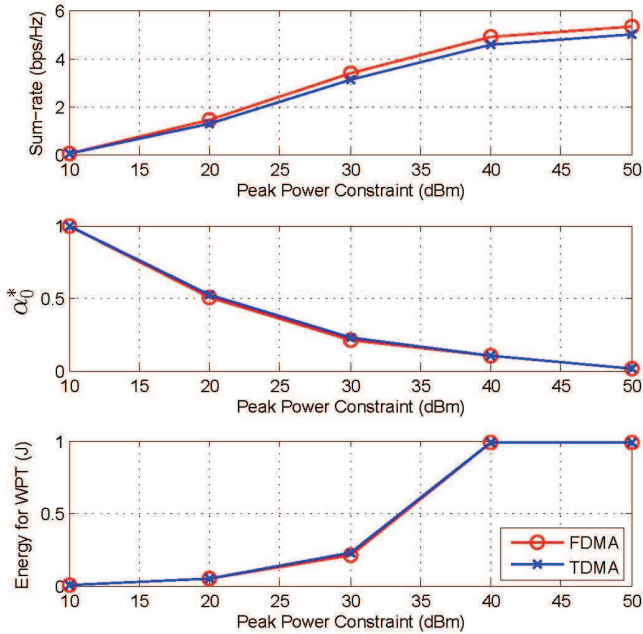}
\vspace{-0.1cm}
\caption{System performance  versus peak power constraint $P_{\rm{peak}}$, where $N=64$ and $P=30$ dBm.}
\label{optimal_a0_value}
\end{centering}
\vspace{-0.1cm}
\end{figure}

Fig. \ref{optimal_a0_value} illustrates  system performance  versus $P_{\rm{peak}}$  with  the fixed total transmit power $P=30$  dBm.
First, it can be observed that the sum-rate increases with $P_{\rm{peak}}$ and reaches a plateau when $P_{\rm peak}>40$ dBm. This is because that the total available power $P$ is fixed and thus the sum-rate must be bounded even if $P_{\rm peak}$ becomes sufficiently large.
Second, for both schemes, we can see that the optimal $\alpha_0^*$ decreases with $P_{\rm peak}$. This is because that less time for WPT is needed to obtain the harvested energy requirement with a larger $P_{\rm peak}$.
In addition, we can also observe that the energy for WPT increases with $P_{\rm peak}$ and then remains fixed for both schemes, where almost all the available energy is used for WPT. This is because that the optimal WPT energy is under the peak power constraint, which becomes infeasible when $P_{\rm peak}$ becomes sufficiently large.

\begin{figure}[t]
\begin{centering}
\includegraphics[scale=0.65]{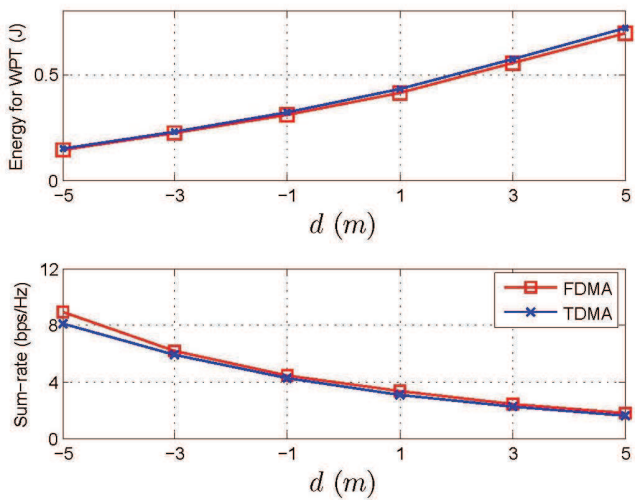}
\vspace{-0.1cm}
\caption{Optimal energy for WPT and sum-rate versus the distance $d_1$, where $N=64$, $P=30$ dBm and $P_{\rm peak}=2P.$}
\label{distance_A_compare_TDMA_FDMA}
\end{centering}
\vspace{-0.1cm}
\end{figure}

Fig. \ref{distance_A_compare_TDMA_FDMA}  examines the effect of the relay position on the energy for WPT and the sum-rate, where the total transmit power is set to be 30 dBm and the HRN moves along with the x-axis from $d=-5$ m to $d=5$ m.
First, we can observe that the energy allocated for WPT is increasing with  $d$ for both cases. This may be because that the channel gains for WPT become worse with a longer distance from sources to HRN, which requires more energy at the HRN allocated to WPT.
Besides, the sum-rate for both schemes is observed to decrease with $d$,  which is different from the traditional relaying systems.
On the one hand, due to the doubly distance-dependent signal attenuation for both WPT and the first hop of WIT, the rate of the first hop cannot be improved though the WPT energy becomes larger.
On the other hand, since the energy for WIT is decreasing with $d$ due to the larger WPT energy, the rate of the second hop is also bottlenecked.  As a result, the HRN should be located in proximity to the sources.

\begin{figure}[t]
\begin{centering}
\includegraphics[scale=0.65]{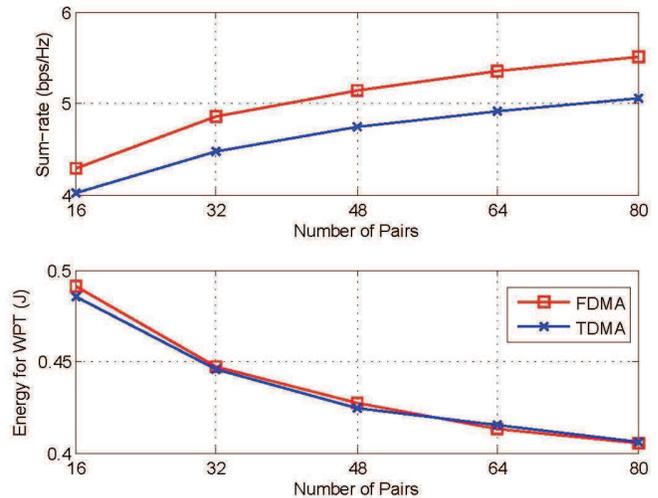}
\vspace{-0.1cm}
\caption{Sum-rate and optimal energy for WPT versus the number of pairs, where $N=64$, $P=30$ dBm and $P_{\rm peak}=2P.$}
\label{Kcompare_TDMA_FDMA}
\end{centering}
\vspace{-0.1cm}
\end{figure}

Fig. \ref{Kcompare_TDMA_FDMA} depicts the sum-rate and the optimal energy for WPT versus the number of pairs $K$ with $P = 30$ dBm.
First, it can be observed that the FDMA scheme achieves higher sum-rate compared to the TDMA scheme.
Besides, we can see that the sum-rate for both schemes increases with the number of pairs and the rate tends to be saturated due to the limited bandwidth and transmission power of the system.
In addition, it can be observed that the energy for WPT decreases with the increasing number of pairs, this may be because that  more pairs can harvest energy from the same energy signals broadcasted by the HRN, which results in  a higher energy efficiency.

%

\section{Conclusions}

This paper studied a new  R-WPCN based on TDMA and FDMA for multiple source-destination pairs with the help of a HRN which acts as both roles of an energy transmitter and an information forwarder.
By considering  the ``charge-then-forward" protocol at the HRN, we studied the sum-rate maximization problem for both TDMA and FDMA cases. For the TDMA case, we proposed a global optimal solution, while for the FDMA case, we designed an asymptotically optimal solution.
To tradeoff the performance and complexity, the suboptimal algorithms for both cases were also proposed.
Extensive simulations showed that our proposed algorithms significantly outperform the conventional schemes.

The following directions  can be considered for possible future works.  First, the HRN is equipped with large-scale antenna array (or massive MIMO). Specifically, massive MIMO can generate concentrated energy beams to power wireless nodes and thus deal with the challenge of long-distance WPT. Second, multiple HRNs with full-duplex could be considered to improve spectral efficiency, where distributed coordinated beamforming at the HRNs for both WPT and WIT will be interesting. Third, it may also consider that the direct link exists between the sources and destinations.  In this case the resource allocation schemes will be largely different. Last, it will be interesting to extend our work to the multiple frames, where sources can accumulate energy and then transmit information over different frames, and the HRN can dynamically allocate its resources over different frames.

\section*{Acknowledgement}
The authors would like to thank the anonymous reviewers. Their valuable suggestions greatly improved the quality of this paper.

\appendices

\section{Proof of Proposition \ref{TDMA_peak_fty} }

It can be proved by contradiction as follows:
%
%
We denote the optimal solution of Problem $\rm (P1')$ as $\{\alpha_k^*\}_{k=0}^{K}$, $\{s_k^*\}_{k=0}^{K}$ and $\{m_k^*\}_{k=1}^{K}$.  Suppose that $s_0^* < \alpha_0^* P_{\rm peak}$, i.e., $p_0^*<P_{\rm peak}$. Then,  we  consider the following solution $s_0^*$ and $\{\tilde{\alpha}_k\}_{k=0}^{K}$, where $s_0^* = \tilde{\alpha}_0 P_{\rm peak}$, i.e., $\tilde{p}_0 = P_{\rm peak}$.
Since $p_0^* < \tilde{p}_0$, we have $\alpha_0^* > \tilde{\alpha}_0$ with the same optimal $s_0^*$. From \eqref{alpha_s_TDMA} and \eqref{alpha_k_peak}, we can obtain that $\tilde{\alpha}_k \geq \alpha_k^*$ for $ k=1,\cdots,K$.
Moreover,  according to constraint  \eqref{C_rate2_TDMA}, as the objective function of Problem $\rm (P1')$ is an increasing function of $\{\alpha_k\}_{k=1}^{K}$ with given $s_k^*$ and $m_k^*$,  the case $\{\tilde{\alpha}_k\}_{k=0}^{K}$ achieves higher sum-rate than the case $\{\alpha_k^*\}_{k=0}^{K}$.
Thus, this contradicts with the assumption that the $\{\alpha_k^*\}_{k=0}^{K}$ is the optimal solution.
Therefore, the optimal solution must be given by $\{\tilde{\alpha}_k\}_{k=0}^{K}$ and $\tilde{p}_0 = s_0^*/\tilde{\alpha}_0 = P_{\rm peak}$, which completes the proof.

\section{Proof of Proposition \ref{sub_TDMA}}

%

We can easily prove that Problem \eqref{sumrate_first_TDMA}  is convex with given $\alpha_0$, which can be optimally solved by the Lagrangian dual method.
%
%
The Lagrangian of Problem \eqref{sumrate_first_TDMA}  is given by
\begin{align}
\mathcal{L}(\boldsymbol{\alpha},\lambda) = \sum\limits_{k=1}^K \frac{\alpha_k}{2}\log_2\left(1+\frac{2A_k }{\alpha_k\sigma^2}\right) + \lambda(1-\sum_{k=0}^{K}\alpha_k),\nonumber
\end{align}
where $\lambda\geq 0$ denotes the Lagrangian multiplier associated with the constraint \eqref{alpha_k_TDMA}. Since the problem is convex, we can find its optimal solution by using Karush-Kuhn-Tucker (KKT) conditions. With given $\alpha_0$, let us denote the primary and dual optimality values of Problem \eqref{sumrate_first_TDMA}  as $\{\alpha_k^*\}_{k=1}^{K}$ and $\lambda^*$. By differentiating $\mathcal{L}(\boldsymbol{\alpha},\lambda)$ with respect to $\alpha_k$ and using KKT stationary conditions, we obtain
\begin{align} \label{daoshu_x}
\frac{1}{2}\log_2(1+2x_k) - \frac{1}{2\ln2} + \frac{1}{2\ln2(1+2x_k)}=\lambda^*,
\end{align}
where $x_k = \frac{A_k }{\alpha_k^*\sigma^2},k=1,\cdots,K.$
To make \eqref{daoshu_x} more clearly, we denote $y_k = \frac{1}{1+2x_k}$, then \eqref{daoshu_x} is equivalent to
\begin{align} \label{daoshu_y}
\log_2y_k = \frac{\ln y_k}{\ln2}= \frac{y_k}{\ln2}-\frac{1}{\ln2}-2\lambda^*.
\end{align}
Thus, we have $-y_k - \delta = -\ln y_k = \ln\frac{1}{y_k}$, where $\delta = -1-2\ln2 \lambda^*$. Then, it is easy for us to get that $e^{-y_k-\delta}=\frac{1}{y_k}$, which is equivalent to $-y_k e^{-y_k}=-e^\delta$. As a result, the solution of \eqref{daoshu_y} is given by $y_k = -W(-e^\delta)$.
Finally, due to $y_k = \frac{1}{1+2x_k}$ and $x_k =\frac{ A_k }{\alpha_k^* \sigma^2}$, we can obtain that the solution of $\frac{\partial\mathcal{L}_k}{\partial\alpha_k}=0$  is given by
\begin{align} \label{TDMA_timeW}
\alpha_k^* = \frac{-2 A_k   W(-e^\delta)}{\sigma^2(1+W(-e^\delta))}, \quad \forall k =1,\cdots,K.
\end{align}

From \eqref{TDMA_timeW}, we can find that $\alpha_k^*,k=1,\cdots,K$ is proportional to $A_k$. Moreover, it can be easily verified that $\sum_{k=1}^{k=K}\alpha_k^*=1-\alpha_0$ must hold for Problem \eqref{sumrate_first_TDMA} with given $\alpha_0$. Thus the optimal $\{\alpha_k^*\}_{k=1}^{K}$ with given $\alpha_0$ is thus given by
\begin{align} \label{alpha_sub}
\alpha_k^* = \frac{ A_k}{\sum_{k=1}^{K} A_k } (1-\alpha_0), \quad \forall k=1\cdots,K.
\end{align}
By plugging \eqref{alpha_sub} into \eqref{q_k+}, then the optimal $\{q_k^*\}_{k=1}^{K}$ with given $\alpha_0$ is given by
\begin{align}
q_k^* = \frac{2}{(1-\alpha_0)h_{1,k}}\sum_{k=1}^{K} A_k, \quad \forall k=1\cdots,K.
\end{align}
The proof is thus completed.

\bibliography{Reference}

%
%
\end{document}